\documentclass[useAMS,usenatbib,usegraphicx]{mn2e}
\DeclareMathVersion{bold}
\title[Scalar Statistics]%
{Scalar quantities as detectors of non-Gaussianity on CMB maps}
\author[Monteser\'\i n et al]{C. Monteser\'\i n$^{1,2}$ R.B. Barreiro$^{1,2}$, E. Mart\'\i nez-Gonz\'alez$^1$ and J.L. Sanz$^1$\\ 
$^1$ Instituto de F\'\i sica de Cantabria, CSIC-Universidad de Cantabria,
Avda. de los Castros s/n, 39005 Santander, Spain\\
$^2$ Dpto. de F\'\i sica Moderna, Universidad de Cantabria, Avda. de
los Castros s/n, 39005 Santander, Spain \\
}

\begin{document}
\maketitle

\begin{abstract}
We study the power of several scalar quantities constructed on the
sphere (presented in Monteser\'\i n et al.) to detect non-Gaussianity
on the temperature distribution of the cosmic microwave background
(CMB). The test has been performed using non-Gaussian CMB simulations
with injected skewness or kurtosis generated through the Edgeworth
expansion. We have also taken into account in the analysis the effect
of anisotropic noise and the presence of a Galactic mask. We find that
the best scalars to detect an excess of skewness in the simulations
are the derivative of the gradient, the fractional isotropy, the
Laplacian and the shape index. For the kurtosis case, the fractional
anisotropy, the Laplacian and the determinant are the quantities that
perform better.
\end{abstract}


\section{Introduction}

The cosmic microwave background (CMB) is currently one of the most
powerful tools of cosmology. Its study provides us with essential
information about the origin and evolution of the Universe. In
particular, a key issue is whether the CMB temperature fluctuations
follow a Gaussian distribution, as predicted by the standard
inflationary model. A detection of intrinsic non-Gaussianity in the
CMB would be a hint of new physics (See~\citealt{bart04} or~\citealt{tur90} 
and references therein) and therefore would have far
reaching consequences on our current knowledge of the Universe. In
addition secondary anisotropies (such as the Sunyaev-Zeldovich
effects, or the Rees-Sciama effect), contaminant astrophysical
emissions (coming from Galactic and extragalactic foregrounds) and
systematics can also leave non-Gaussian imprints in the observed
microwave sky. Therefore, it is crucial to perform a careful study of
any possible detected non-Gaussianity in order to understand its
origin.

In the last years there has been a large number of experiments
measuring the CMB anisotropies and polarization
(DASI,~\citealt{hal02}; VSA,~\citealt{gra03}; CBI,~\citealt{mas03};
ACBAR,~\citealt{kuo04}; Archeops,~\citealt{beno03}). In particular,
the NASA WMAP satellite has provided with high quality observations of
the whole sky (\citealt{benn03a}). Many works have studied the
Gaussianity of the WMAP data, including the use of wavelet tools
(\citealt{vie04},~\citealt{muk04},~\citealt{cru05},~\citealt{cru06a},
~\citealt{cru06b},~\citealt{mce05} and~\citealt*{cay05}), harmonic techniques
(\citealt{kom03},~\citealt{chi03},~\citealt{col04},~\citealt*{nas04},
~\citealt{mag04}), and real space analyses
(\citealt{par04},~\citealt{eri04a},
~\citealt{han04},~\citealt{lar04},~\citealt{lar05},~\citealt{cabe05}),
finding in some cases deviations from Gaussianity whose origin is
uncertain.
In this paper, we are interested in the study of the Gaussianity of
the CMB using a set of scalar quantities of the temperature field,
such as the Laplacian, the shape index or the Gaussian
curvature. Previous works on related methods include the study of
maxima properties (\citealt{bond87},~
\citealt{barr97},~\citealt{gur97},~\citealt*{barr01}), the fraction of
lake and hill points (\citealt*{dor03},~\citealt{cabe05}) or the
skeleton lenght (\citealt*{nov06},~\citealt{eri04b}).

Some previous works have also compared the performance of 
different statistics to detect non-Gaussianity in the CMB using a
reference set of non-Gaussian simulations. For instance, \cite{hob99}
compared the capability of wavelet techniques and Minkowski
functionals as non-Gaussianity detectors using simulated maps of the
Kaiser-Stebbins effect. The comparative performance of the Spherical
Haar Wavelet and the Spherical Mexican Hat Wavelet has been tested on
non-Gaussian simulations generated through the Edgeworth expansion
\citep{mar02}. \cite{agh03} studied the relative performance of
wavelet and Fourier based methods by applying them to three different
sets of non-Gaussian maps that were representative of possible
non-Gaussian signatures present in the CMB. Three different approaches
in pixel, harmonic and wavelet space were tested by \cite{cabe04} on
simulations of the quadratic potential model and on Gaussian maps
contaminated by point sources. \cite{jin05} have compared
theoretically and experimentally the performance of the kurtosis to
the Higher Criticism and MAX statistics in the wavelet and curvelet
spaces.

In a previous work,~\cite{mon05} (hereafter M05) studied the
probability distribution of a series of scalar quantities assuming
that the CMB is a homogeneous and isotropic Gaussian field and
including also the effect of anisotropic noise and the presence of a
mask. These quantities were proposed as promising tools for the
analysis of the temperature distribution of the CMB. In the present
paper, we study the performance of these scalars (as well
as three new ones) to detect non-Gaussianity in the CMB. In
particular, we have tested the method using non-Gaussian simulations
with skewness and kurtosis injected through the Edgeworth expansion
(\citealt{mar02}).  Although Edgeworth simulations do not correspond
to a realistic model, it is expected that physical motivated models,
as well as contaminant emissions, will introduce a certain level of
skewness and/or kurtosis in the CMB distribution. As an example, the
quadratic potential model would introduce skewness in the CMB
(\citealt{kom01}), the cosmic strings would generate kurtosis
(\citealt*{hob99},~\citealt{barh01}) whereas residual point sources
would introduce both (\citealt{arg06}).  Therefore the use of
non-Gaussian simulations constructed through the Edgeworth expansion,
where a certain level of skewness or kurtosis is injected, provides us
with an interesting tool in order to test the type of deviations from
Gaussianity that one would expect in realistic data.

The article is organised as follows. In $\S$2 we briefly summarise the
work of M05, describing the different considered scalars 
(including three new quantities)
and presenting some theoretical results for the case of a Gaussian
field. Our simulations are described in $\S$3 whereas $\S$4 presents
the results of our Gaussianity test. Section $\S$5 summarises our main
conclusiones. Finally, we include a series of appendices with the description
of the three new quantities as well as analytical expressions for the 
cumulative functions of some of the considered scalars.

\section{Scalars on the sphere}
\label{sec:scalars}
The temperature anisotropies of the CMB are usually described as a
2-dimensional field T$(\theta, \phi)$ on the sphere. The first and
second derivatives of the field encode very interesting information
about the fluctuations. In particular, different quantities which are
scalars under a change of the coordinate system (i.e. regular general
transformation $(x_1,x_2) \to (x_1',x_2'): s'(x_1',x_2') = 
s(x_1,x_2))$ can be constructed from the first and second covariant
derivatives of the field, which can be useful to perform Gaussian
studies of the CMB. Since a field can deviate from Gaussianity in an
infinite number of ways, it is a non-trivial task to design a unique
set of scalars that is optimal to detect any possible type of
non-Gaussianity. Therefore our aim is to study the performance of a
(relatively large) number of scalars to detect some generic types of
non-Gaussianity (non-zero skewness and kurtosis).
In particular, M05 studied the probability 
density function of different scalars for a homogeneous and isotropic 
Gaussian field. In this section we briefly describe these quantities 
(for a more detailed description, see M05) and introduce 
three new scalars (curvedness, fractional isotropy and fractional 
anisotropy).

Using first derivatives we can construct the square of the modulus of
the gradient $g$, which provides information on how the temperature
varies spatially, and is given by:
\begin{equation}
g = T^{,i}T_{,i} ~.
\end{equation}
Other scalars can be constructed using only second derivatives.  These
scalars are directly related to the eigenvalues $\lambda_i$ of the negative
Hessian matrix of the field, $\left[-T_{;ij}\right]$. In particular
the eigenvalues can be written as a function of the covariant second
derivatives in the following way:
\begin{equation}
\lambda_{1}=-\frac{1}{2} \left[ \left(T_{\ \ i}^{;i}\right) - 
\sqrt{\left(T_{\ \ i}^{;i}\right)^2 - 2 \left(T_{\ \ i}^{;i}T_{\ \ j}^{;j} - T_{\ \ i}^{;j} T_{\ \ j}^{;i} ~,
\right)} \right] 
\end{equation}
\begin{equation}
\lambda_{2} = -\frac{1}{2} \left[ \left(T_{\ \ i}^{;i}\right) + 
\sqrt{\left(T_{\ \ i}^{;i}\right)^2 - 2 \left(T_{\ \ i}^{;i}T_{\ \ j}^{;j} - 
T_{\ \ i}^{;j} T_{\ \ j}^{;i} \right)} \right] ~, 
\end{equation}
Combining these three quantities, we can construct other scalars that
enhance different features or properties of the analysed temperature
field. An interesting scalar is given by the trace of the Hessian
matrix (the Laplacian) $\lambda_{+}$:
\begin{equation}
\lambda_{+} = -\lambda_{1}-\lambda_{2} ~.
\end{equation}
The distortion $\lambda_{-}$ and the shear $y$ are related to the
difference between eigenvalues:
\begin{equation}
\lambda_{-} = \lambda_{1}-\lambda_{2} ~,
\end{equation}
\begin{equation}
y = \frac{1}{4} (\lambda_{1}-\lambda_{2})^2 ~. 
\end{equation}
While these scalars describe the asymmetry of the field through the
difference of the eigenvalues, other scalars express a similar
information using their ratio. The ellipticity $e$ and its bounded
construction the shape index $\iota$ are dimensionless scalars given
by:
\begin{eqnarray}
e = \frac{\lambda_{1} - \lambda_{2}}{2(\lambda_{1} + \lambda_{2})} ~,
\end{eqnarray} 
\begin{equation}
\iota = \frac{2}{\pi} \arctan{\left(-\frac{1}{2e}\right)}
\end{equation}
Other scalars, related to the curvature of the field, are the
determinant $d$ of the Hessian matrix, the Gaussian curvature
$\kappa_G$ and the curvedness $c$ (see appendix \ref{ap:curvedness_ap}):
\begin{equation}
d = \lambda_{1} \lambda_{2} ~,
\end{equation} 
\begin{equation}
\kappa_{G} = \frac{\lambda_{1} \lambda_{2}}{\left(1 + g \right)^{2}}  ~,
\end{equation}
\begin{equation}
c = \frac{1}{2} \sqrt{\lambda_{1}^{2} + \lambda_{2}^{2} } ~.
\end{equation}
We can also construct new scalars by combining two of the previous
scalars.  Examples of these are the fractional isotropy $f_{i}$ and
the fractional anisotropy $f_a$, which correspond, respectively, to
the Laplacian and the distortion weighted by the curvedness (see
appendices \ref{ap:frac_aniso_ap} and \ref{ap:frac_iso_ap}):
\begin{equation}
f_{i} = \frac{1}{\sqrt{2}} \frac{\lambda_{1}+\lambda_{2}}{\sqrt{\lambda_{1}^{2}+\lambda_{2}^{2}}} ~,
\end{equation}
\begin{equation}
f_{a} = \frac{1}{\sqrt{2}} \frac{\lambda_{1}-\lambda_{2}}{\sqrt{\lambda_{1}^{2}+\lambda_{2}^{2}}} ~.
\end{equation}

Finally, we have also considered the derivative of the square of the
gradient modulus, $D_{g}$:
\begin{equation}
D_{g} = T^{;ij}T_{,i}T_{,j} ~,
\label{eq:Dg}
\end{equation}

Assuming that T$(\theta, \phi)$ is a homogeneous and isotropic
Gaussian random field, we can derive analytical or semi-analytical
expressions for the probability density function (pdf) of the previous
scalars (see M05 and appendices \ref{ap:curvedness_ap} to
\ref{ap:frac_iso_ap}). In this case, the pdf of the different scalars
is completely determined by the power spectrum, $C_{\ell}$ of the
original field T. This dependence appears in the distribution function
of the scalars through the moments of the initial field, $\sigma_i$:
\begin{equation}
\sigma_{i}^2 = \sum_{\ell} C_{\ell}
\frac{2\ell+1}{4\pi}\left[\ell\left(\ell+1\right)\right]^{i} ~.
\label{eq:momentos}
\end{equation}
However, as shown in M05, it is possible to remove the dependence of
the pdf of the scalars on the power spectrum of the field by
constructing new quantities that we have called {\it normalised
scalars}. Therefore, for the Gaussian case, a given normalised scalar
will have the same distribution function independently of the power
spectrum of the field. The definitions of the normalised scalars -- as
a function of the ordinary scalars -- as well as their pdf's for a
homogeneous and isotropic Gaussian field are given in table
\ref{tab:norm_scalars} (for completeness we also include the
normalised temperature in the table). In addition to its greater
simplicity, these new quantities allow us to deal in a straightforward
way with the presence of anisotropic noise (see below).
\begin{table*}
\begin{center}
    \begin{tabular}{ccccc}
      \hline
       Normalised scalar & Notation &  Definition & Domain&  pdf \\
      \hline
temperature & $\tilde{T}$& $\frac{T}{\sigma_{0}}$ & $(-\infty,\infty)$ & $p(\tilde{T})=\frac{1}{\sqrt{2\pi}} e^{\frac{-\tilde{T}^{2}}{2}}$  \\
      \hline
greatest eigenvalue & $\tilde{\lambda}_{1}$& $\frac{1}{2} \left( \tilde{\lambda}_{+}+ \tilde{\lambda}_{-} \right)$ & $(\tilde{\lambda}_{2},\infty)$ & $p(\tilde{\lambda}_{1})= \frac{4}{3\sqrt{2\pi}} e^{-2\tilde{\lambda}_{1}^{2}} \left( 1 + \sqrt{\frac{2\pi}{3}} \tilde{\lambda}_{1} e^{2 \frac{\tilde{\lambda}_{1}^{2}}{3}} \left[ 1 + erf\left(\sqrt{\frac{2}{3}} \tilde{\lambda}_{1}\right) \right] \right)$   \\
      \hline
lowest eigenvalue & $\tilde{\lambda}_{2}$& $\frac{1}{2} \left( \tilde{\lambda}_{+} - \tilde{\lambda}_{-} \right)$ & $(-\infty,\tilde{\lambda}_{1})$ & $p(\tilde{\lambda}_{2})= \frac{4}{3\sqrt{2\pi}} e^{-2\tilde{\lambda}_{1}^{2}} \left( 1 - \sqrt{\frac{2\pi}{3}} \tilde{\lambda}_{1} e^{2 \frac{\tilde{\lambda}_{1}^{2}}{3}} \left[ 1 - erf\left(\sqrt{\frac{2}{3}} \tilde{\lambda}_{1}\right) \right] \right)$ \\
      \hline
Laplacian & $\tilde{\lambda}_{+}$& $-\frac{\lambda_{+}}{\sigma_{2}}$ & $(-\infty,\infty)$ & $p(\tilde{\lambda}_{+})=\frac{1}{\sqrt{2\pi}} e^{\frac{-\tilde{\lambda}_{+}^{2}}{2}}$  \\
      \hline
determinant & $\tilde{d}$ & $\tilde{\lambda}_{1} \tilde{\lambda}_{2}$ & $(-\infty,\infty)$ & $p(\tilde{d}) = \left\{ \begin{array}{ll} \frac{4}{\sqrt{3}} e^{4\tilde{d}} & \tilde{d} < 0 \\
\frac{4}{\sqrt{3}} e^{4\tilde{d}} \left[1 - erf\left( \sqrt{6\tilde{d}}\right)    \right] &  \tilde{d} > 0 \\
\end{array} \right .$   \\
      \hline
shear & $\tilde{y}$ & $\frac{y}{\sigma_{2}^{2}-2\sigma_{1}^{2}}$ & $(0,\infty)$ & $p(\tilde{y})=4e^{-4\tilde{y}}$ \\
      \hline
distortion & $\tilde{\lambda}_{-}$& $\frac{\lambda_{-}}{\sqrt{\sigma_{2}^{2}-2\sigma_{1}^{2}}}$ & $(0,\infty)$ & $p(\tilde{\lambda}_{-})=2 \tilde{\lambda}_{-} e^{-\tilde{\lambda}_{-}^{2}}$               \\
      \hline
ellipticity & $\tilde{e}$ & $\frac{e\sigma_{2}}{\sqrt{\sigma_{2}^{2}-2\sigma_{1}^{2}}}$ & $(-\infty,\infty)$ &  $p(\tilde{e})=4 \vert \tilde{e}\vert \left( 1+8\tilde{e}^2 \right)^{-\frac{3}{2}}$                             \\ 
      \hline 
shape index & $\tilde{\iota}$& $\frac{2}{\pi} \arctan{\left(-\frac{1}{2\tilde{e}}\right)}$ & $(-1,1)$ & $p(\tilde{\iota})= \frac{\pi}{2} \frac{\vert \cos{\left(\frac{\pi}{2} \tilde{\iota} \right)}\vert }{\vert \sin^{3}{\left(\frac{\pi}{2} \tilde{\iota}\right)} \vert} \left[ 1 + 2 \cot^{2}{\left(\frac{\pi}{2} \tilde{\iota}\right)}\right]^{-\frac{3}{2}}$                      \\
      \hline
fractional anisotropy & $\tilde{f}_{a}$ & $\frac{1}{\sqrt{2}} \frac{\tilde{\lambda}_{1}-\tilde{\lambda}_{2}}{\sqrt{\tilde{\lambda}_{1}^{2} + \tilde{\lambda}_{2}^{2}}}$ & $(0,1)$ & $p(\tilde{f}_{a})=\frac{2\tilde{f}_{a}}{\left( 1-\tilde{f}_{a}^{2}\right)^\frac{1}{2} \left( 1+\tilde{f}_{a}^{2}\right)^\frac{3}{2}}$   \\
      \hline
fractional isotropy & $\tilde{f}_{i}$ & $\frac{1}{\sqrt{2}} \frac{\tilde{\lambda}_{1}+\tilde{\lambda}_{2}}{\sqrt{\tilde{\lambda}_{1}^{2}+\tilde{\lambda}_{2}^{2}}}$ & $(-1,1)$ & $p(\tilde{f}_{i})=\frac{1}{\left(2-\tilde{f}_{i}^{2} \right)^{\frac{3}{2}}}$   \\
      \hline
gradient & $\tilde{g}$& $\frac{g}{\sigma_{1}^{2}}$ & $(0,\infty)$ & $p(\tilde{g})=e^{-\tilde{g}}$ \\
      \hline
derivative of gradient & $\tilde{D}_{g}$& $\frac{D_{g}}{\frac{1}{2} \sigma_{1}^{2}\sqrt{3\sigma_{2}^{2}-2\sigma_{1}^{2}}}$ & $(-\infty,\infty)$ & $p(\tilde{D}_{g})=\frac{2}{\sqrt{\pi}} \int_{0}^{\infty}{e^{-z^{2}-\frac{\tilde{D}_{g}^2}{z^{4}}} \frac{\it{d}z}{z}}$ \\
      \hline 
Gaussian curvature & $\tilde{\kappa}_{G}$ & $\frac{\tilde{d}}{\left( 1 + \tilde{g}\right)^{2}} $ & $(-\infty,\infty)$  & $p(\tilde{\kappa}_{G}) = \left\{ \begin{array}{ll}
-\int_{-\infty}^{\tilde{\kappa}_{G}}\frac{2e}{\tilde{\kappa}_{G}} \sqrt{\frac{z}{3\tilde{\kappa}_{G}}}  e^{-\sqrt{\frac{z}{\tilde{\kappa}_{G}}}} e^{4z} dz & \tilde{\kappa}_{G} < 0 \\
\int_{\tilde{\kappa}_{G}}^{\infty}\frac{2e}{\tilde{\kappa}_{G}} \sqrt{\frac{z}{3\tilde{\kappa}_{G}}}  e^{-\sqrt{\frac{z}{\tilde{\kappa}_{G}}}} e^{4z} \left[ 1 -  erf \left( \sqrt{6z}\right)\right] dz & \tilde{\kappa}_{G} > 0 \\
\end{array}
\right .$       \\
      \hline 
Curvedness & $\tilde{c}$ & $\frac{1}{2} \sqrt{\tilde{\lambda}_{+}^{2} +\tilde{\lambda}_{-}^{2}}$ & $(0, \infty)$ & $p(\tilde{c})=\frac{16}{\sqrt{\pi}} \vert \tilde{c}\vert e^{-4 \tilde{c}^{2}} D \left(\sqrt{2} \tilde{c} \right)$ \\
      \hline 
    \end{tabular}
    \caption{\label{tab:norm_scalars} Definition of the normalised
scalars as a function of the ordinary ones. The pdf's correspond to
a Gaussian initial field and are
independent of its power spectrum. Note that the values of $\sigma_{i}$
may be pixel-dependent.}
    \end{center}
  \end{table*}

We have to point out that many of the considered normalised scalars
have some level of correlation between each other. For a Gaussian
field the maximum number of uncorrelated scalars is three, for scalars
constructed from the first and second derivatives. In particular, the
square of the modulus of the gradient is uncorrelated with all the
other scalars. Therefore a set of three uncorrelated scalars will be
formed by $g$ and two other scalars, whose choice is not unique.  All
the information about correlations between the normalised scalars is
given in table 2 of M05 and table \ref{tab:correl_c_fa_fi} of this
work.

In order to construct the normalised scalars of a field, a necessary
step is the calculation of the moments $\sigma_0$, $\sigma_1$ and
$\sigma_2$ given by equation (\ref{eq:momentos}) through the
$C_\ell$'s. For a noiseless CMB map this is trivial provided that we
know its power spectrum. However, in the presence of anisotropic
Gaussian noise and a mask some further considerations need to be
made. Given that the noise and the mask introduce discontinuities in
the CMB map, which represents a problem for the calculation of the
derivatives, we first filter the considered map with a Gaussian beam
of dispersion $\sigma_g$ (which is chosen to be equal to the
dispersion of the beam of the observed CMB map). The
anisotropic noise is characterized by a different dispersion at each
pixel, $\sigma_{n}(\vec{x})$ (with $\vec{x}$ the unity vector on
the sphere in the direction of observation). In order to construct the
normalised scalars we need to calculate the moments of the field
taking into account the different noise in each pixel. This is done
using a {\it pixel dependent power spectrum} $H_{\ell}(\vec{x})$ (see
M05):
\begin{equation}
H_{\ell}(\vec{x}) = \left[C_{\ell}^{s} e^{-\ell \left(\ell +1 \right)
\sigma_{g}^{2}} + \frac{4 \pi
\sigma_{n}^{2}(\vec{x})}{N_{pix}}\right] e^{-\ell \left(\ell +1
\right) \sigma_{g}^{2}} 
\label{eq:spectrum_aninoise}
\end{equation}
where $N_{pix}$ is the total number of pixels of the map.
$H_{\ell}(\vec{x})$, for fixed $\vec{x}$, corresponds to the power
spectrum of a map containing the filtered CMB signal plus isotropic
noise with dispersion $\sigma_{n}(\vec{x})$ and filtered again with a
Gaussian beam of dispersion $\sigma_g$.  Note that, due to this extra
smoothing, the dispersion of the noise in a given pixel will depend
not only on the noise level on that position but also on the
$\sigma_n$ of its neighbouring pixels. However provided that
$\sigma_n(\vec{x})$ varies smoothly, as it is usually the case, this
effect is small. 

Therefore, using $H_{\ell}(\vec{x})$, we can obtain the moments of the
field $\sigma_i(\vec{x})$ at each pixel and, from them, construct the
normalised scalars by introducing these pixel dependent moments on
their corresponding definitions. The normalised scalars constructed
in this way for a Gaussian initial field will follow the pdf's given
in table~\ref{tab:norm_scalars}, which are independent of the
underlying power spectrum and the level of anisotropic noise.

We would like to point out that if the dispersion of the noise
does not vary smoothly along the map, equation
(\ref{eq:spectrum_aninoise}) may not be a sufficiently good
approximation to obtain $\sigma_i(\vec{x})$. In this case, the
contribution of the noise to the moments of the field can still be
obtained using simulations. Note that since signal and noise are
independent, their contributions to $\sigma_i(\vec{x})$ can be
calculated separately and then simply added together
(quadratically). In order to obtain numerically the contribution of
the noise to $\sigma_i(\vec{x})$, we need to generate a large number
(1000) of noise simulations, that are smoothed with a Gaussian beam of
dispersion $\sigma_g$. We obtain then the dispersion of the noise
$\sigma_0$ in the smoothed map at a given pixel, as the dispersion of
the values of the 1000 simulations at that pixel. To obtain
$\sigma_1(\vec{x})$ and $\sigma_2(\vec{x})$, we need to construct two
ordinary scalars, the square of the modulus of the gradient $g$ and
the Laplacian $\lambda_+$, corresponding to the previous smoothed
noise simulations. As shown in M05, the mean value of $g$ at each
pixel is given by $\sigma_1^2(\vec{x})$ whereas $\sigma_2(\vec{x})$
corresponds to the dispersion of $\lambda_+$ at each
position. Therefore, by estimating these quantities at each pixel from
the ordinary gradient and Laplacian of the smoothed noise simulations,
we obtain the contribution of the noise to the moments of the field.

By comparing the estimated values of $\sigma_i(\vec{x})$ using the
numerical and analytical (approximated) methods, we can
quantify which is the error introduced in these quantities when using
the approximated technique. In particular, we have calculated
$\sigma_i(\vec{x})$ using both methods for the example considered in
this work, which corresponds to the level of noise expected in the
Q+V+W combined map of the 4-year WMAP data (see
$\S$\ref{sec:simulations}). We find that the average
error\footnote{This error is defined as $100 \times \left| \sigma_i^n
- \sigma_i^a \right|/\sigma_i^n$, where $\sigma_i^n$ and $\sigma_i^a$
refer to the $i$th moment obtained with the numerical and approximated
approaches repectively} on the calculation of the noise contribution
to $\sigma_i(\vec{x})$ is less than 2 per cent for the three moments,
with maximum errors less than 15 per cent for every pixel. Moreover,
if we consider the error in $\sigma_i(\vec{x})$, including the
contribution of both the signal and the noise, the average error
becomes negligible (less than 0.1 per cent) whereas the maximum error
is less than 1 per cent for every pixel and for the three moments
considered. Therefore, for the considered work, we have used the
simpler approach given by equation (\ref{eq:spectrum_aninoise}) to
estimate the moments $\sigma_i(\vec{x})$ since the approximation works
very well.

Regarding the effect of the mask, the smoothing of the masked map
(with the pixels of the mask set to zero) reduces the discontinuty at
its boundary. However, this smoothing strongly contaminantes the
pixels close to the boundary of the mask, and therefore the scalars
calculated at those positions, which must be removed from the
analysis.  In order to do this, we will consider in our study only
those pixels outside an extended mask that excludes also those pixels
significantly contaminated by the smoothing of the original mask.

This extended mask is obtained following the heuristic approach
proposed by \citep{eri04b}.  First, the original mask is smoothed with
a Gaussian beam of dispersion 3 times the pixel size. Then, all those
pixels above 0.991 are set to 1 (kept from the analysis) whereas those
below this value are set to 0 (excluded from the analysis). Note that
this way of constructing the mask is different and more conservative
than the one presented in M05. In any case we have tested that both
methods provide very similar results, but we have chosen the one of
the present work due to its simplicity.

\section{The simulations} 
\label{sec:simulations}

In order to test the performance of the considered scalars to detect
non-Gaussianity, we have generated different sets of Gaussian and
non-Gaussian CMB simulations. In particular, we have used the
Edgeworth expansion to simulate non-Gaussian CMB maps with a certain
level of skewness or kurtosis. Although the simulations obtained in
this way do not correspond to a particular physical model, they are,
nonetheless, a useful tool to mimic some generic deviations of
Gaussianity expected in realistic data. Indeed many physical motivated
models, as well as contaminant emissions, will produce a certain level
of these higher order moments on the CMB observations such as the
quadratic potential inflationary model (which introduces skewness,
\citealt{kom01}), cosmic strings (which generates kurtosis,
~\citealt{hob99},~\citealt{barh01}) or point source residuals (which produce
both,~\citealt{arg06}).

The Edgeworth expansion of a one-point density function $f(y)$ can be
expressed in terms of the Hermite polynomials (see e.g. \citealt{mar02} 
and references therein). Keeping only the
first terms in the corresponding Hermite polynomials and considering
only the skewness and kurtosis perturbations, we approximate the
distribution function of our non-Gaussian simulations by the following
two equations:
\begin{equation}
f_S \left(z \right) = \frac{e^{-\frac{z^2}{2}}}{\sqrt{2\pi}} \left\{1 + \frac{S}{6} \left[z \left( z^2 - 3 \right) \right]\right\}
\label{eq:Edge_skewness}
\end{equation}
\begin{equation}
f_K\left(z\right) = \frac{e^{-\frac{z^2}{2}}}{\sqrt{2\pi}}\left[1 + \frac{K}{24} \left(z^4 -6 z^2 +3 \right) \right]
\label{eq:Edge_kurtosis}
\end{equation}
where $S$ and $K$ denote skewness and kurtosis, espectively. These
distribution functions are not always well defined, because they can
become negative even for relatively small values of $S$ or $K$. To
avoid this problem we set the function to zero when it becomes
negative and then renormalise it to unit area. Note that for values of
$S$, $K \la$ 1, the zeroes of the function always appear in the tails
of the distribution, so the renormalization value is close to 1.

Our test CMB simulations have been produced with the aid of the
HEALPix package (\citealt{gor05}), using a resolution of
$N_{side}=256$ (which corresponds to a pixel of 13.7
arcminutes). Following \cite{mar02}, we have generated non-Gaussian
Edgeworth simulations as follows. First we produce simulations of
white noise with a distribution given by equations
(\ref{eq:Edge_skewness}) and (\ref{eq:Edge_kurtosis}) for different
values of $S$ and $K$ (considering also the case $S=K=0$, which
corresponds to a Gaussian distribution). These maps are then convolved
with a Gaussian beam of FWHM=23 arcmin. Finally we
renormalize the power spectrum of the resulting map to the desired CMB
power spectrum. In particular, we have used the power spectrum given
by the best-fit model to the 1-year WMAP data (\citealt{sper03})
convolved with a Gaussian beam of FWHM=33 arcmin. The $C_\ell$'s for
this model were generated using CMBFAST (\citealt{sel96}).

Note that as a consequence of the first beam convolution and the
introduction of correlations in the temperature maps, the simulated
field does not longer follow the expressions given by equations
(\ref{eq:Edge_skewness}) and (\ref{eq:Edge_kurtosis}) and, in
particular, the levels of the injected $S$ and $K$ have been
significantly reduced. Nevertheless, the important point is that this
process generates non-Gaussian simulations with the desired power
spectrum and whose final level of $S$ and $K$ can be controlled
through the values of the original injected skewness and kurtosis. The
upper part of table \ref{tab:Edgeworth_SK_levels} gives the mean and
standard deviation of the final skewness and kurtosis obtained from
1000 simulations for different injected values of $S$ and $K$.  Note
that the reduction in the level of $K$ is more important than the one
in $S$.
\begin{table*}
 \begin{center}
  \caption{\label{tab:Edgeworth_SK_levels} Mean and standard deviation
  of the skewness and kurtosis of maps simulated through the Edgeworth
  expansion for different injected values of $S$ and $K$. The results
  have been obtained from 1000 simulations for each considered
  case.}
  \begin{tabular}{c c c c c c c c}
  \hline
&  injected $S$& injected $K$ & $<S>$ & $\sigma_{S}$&
$<K>$ & $\sigma_{K}$& \\
  \hline
 &0.00& 0.00& $ -1.41 \times 10^{-3} $&$ 3.23 \times 10^{-2}$& $-9.02
  \times 10^{-3} $&$ 3.38 \times 10^{-2}$& \\ 
Ideal &0.08& 0.00& $ 1.18 \times 10^{-2} $&$ 3.20 \times 10^{-2}$& $-8.08
  \times 10^{-3} $&$ 3.45 \times 10^{-2}$& \\ 
&0.00& 0.40& $ 8.97 \times 10^{-4} $&$ 3.45 \times 10^{-2}$& $5.82
  \times 10^{-3} $&$ 3.62 \times 10^{-2}$& \\ 
  \hline
 &0.00& 0.00& $ 2.82 \times 10^{-4} $&$ 4.59 \times 10^{-2}$& $-1.20 \times 10^{-2} $&$ 5.48 \times 10^{-2}$& \\
Realistic &0.20& 0.00& $ 2.21 \times 10^{-2} $&$ 4.63 \times 10^{-2}$& $-1.17 \times 10^{-2} $&$ 5.33 \times 10^{-2}$& \\
&0.00& 1.60& $ -1.55 \times 10^{-3} $&$ 4.55 \times 10^{-2}$& $1.78 \times 10^{-2} $&$ 5.21 \times 10^{-2}$& \\
  \hline
  \end{tabular}
 \end{center}
\end{table*}

As a first step to test the power of the scalars to detect
non-Gaussianity we have applied our method to these three sets of
ideal simulations: Gaussian, injected skewness ($S=0.08$) and injected
kurtosis ($K$=0.4) (section~\ref{sec:noiseless}). Note that we have
chosen to inject in the simulations only skewness or kurtosis to
discriminate better the effect of each of these higher moments in the
Gaussianity test. We remark that the particular values of $S$ and $K$
have been selected to allow for a good comparison between the
performance of the different scalars (although consistent results were
obtained using different values of these parameters).

However, in a realistic experiment, the data will be contaminated by
some level of instrumental noise. Moreover, some regions of the sky
may be severely contaminated by different astrophysical emissions and
thus will be excluded from the analysis. Therefore, we have also
analysed a set of realistic simulations that take into account these
problematics (section~\ref{sec:noisy}). In addition to the CMB
signal, our realistic simulations contain anisotropic Gaussian noise
at the level expected in the Q+V+W combined map of the 4-year WMAP
data and has been masked using the Kp0 mask of the WMAP team 
(\citealt{benn03b}). As explained in the previous section, in order to
reduce the discontinuities produce by the noise and the mask, we have
further smoothed the simulations with a Gaussian beam of FWHM=33
arcminutes. Finally, following the procedure explained in
section~\ref{sec:scalars}, we have constructed an extended mask that
excludes from the Gaussian analysis those pixels contaminated by the
original mask due to this extra filtering. Note that whereas the Kp0
mask covers approximately a 23.9$\%$ of the full sky, the extended Kp0
mask used in the Gaussian analysis excludes a total of 31.7$\%$ of the
pixels.

For the realistic case, we have considered also three different sets
of simulations: Gaussian, injected skewness and injected
kurtosis. Obviously, due to the presence of noise and mask, it is more
difficult to distinguish between different models. Therefore, in order
to obtain a meaningful comparison between the power of the different
scalars, we have chosen in this case higher values for the injected
skewness ($S=0.2$) and kurtosis ($K=1.6$). The lower part of table
\ref{tab:Edgeworth_SK_levels} gives the corresponding mean and
dispersion of the final skewness and kurtosis obtained from 1000
realistic simulations.

\section{Results}

In the present section we have tested the performance of the different
normalised scalars\footnote{Note that in this section, even if not
said explicitely, when talking about scalars we will
always refer to the normalised quantities.} to detect generic
deviations from non-Gaussianity using the sets of simulations
previously described. In order to discriminate between the Gaussian
and non-Gaussian cases, we have applied the well-known
Kolmogorov-Smirnov (KS) test (e.g.~\citealt{von64}) for each
scalar. Given a scalar map, the test simply consists on obtaining the
maximum distance between the empirically constructed cumulative
function of the scalar map and the expected theoretical one
corresponding to the null hypothesis (in our case that the underlying
temperature map is Gaussian). In most cases this theoretical
cumulative function can be obtained, either analytically or
numerically, for each scalar from the pdf's given in
table~\ref{tab:norm_scalars} (see appendix~\ref{ap:cum_func_ap} for
the analytical expressions of the cumulative functions of some
scalars). Nonetheless, for two scalars (derivative of the gradient and
Gaussian curvature) with complicated pdf's we found more convenient to
obtain it simply as the average of the empirically constructed
cumulative function of the 1000 Gaussian simulations.

For a given normalised scalar, the procedure is as follows. First we
obtain the scalar map for our sets of Gaussian and non-Gaussian
simulations. Then we construct the probability distribution of KS
distances for each set of simulations. Afterwards we compare the
Gaussian case with the corresponding non-Gaussian one (either ideal or
realistic). The capability of the scalar to discriminate between both
cases is quantified by obtaining the power $p$ of the test at a given
significance level $\alpha$ (see e.g.~\citealt{cow98}). For a
significance level $\alpha$, we reject the null hypothesis if a
simulation has a KS distance, $d_{KS}$, higher than that of a fraction
$1-\alpha$ of the Gaussian simulations. Therefore $\alpha$ defines a
critical value $d_c$ below (above) which we accept (reject) the null
hypothesis.  $p$ is defined as the fraction of simulations of the
alternative hypothesis with values of $d_{KS}$ higher than $d_{c}$.
Therefore, for a fixed value of $\alpha$, a large value of $p$
indicates that there is a small overlap between both probability
distributions and thus the models can be distinguished.  We have
performed this study for all the considered scalars, as well as for
the normalised temperature map, using the ideal and realistic sets of
Gaussian and non-Gaussian simulations described in the previous
section. The power of the scalars to discriminate between the
different cases has then been compared.

\subsection{Ideal case}
\label{sec:noiseless}

\begin{figure*}
 \begin{center} 
  \includegraphics[angle=0,width=15.0cm]{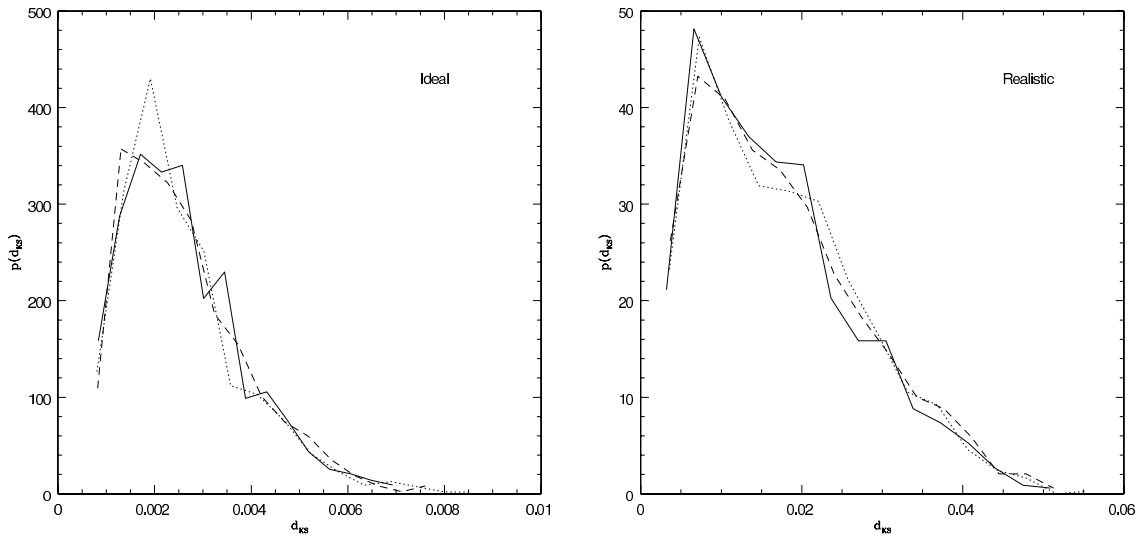}
  \caption{\label{fig:KS_temp_nonoise} Probability distributions of
  the KS distances $p(d_{KS})$ corresponding to the CMB normalised
  temperature maps, each of them obtained from 1000 simulations, for
  the ideal (left panel) and realistic (right panel) cases. The
  different lines correspond to the Gaussian (solid), skewness
  (dashed) and kurtosis (dotted) sets.}
\end{center}
\end{figure*}

\begin{figure*}
 \begin{center}
  \includegraphics[angle=0,width=15.0cm]{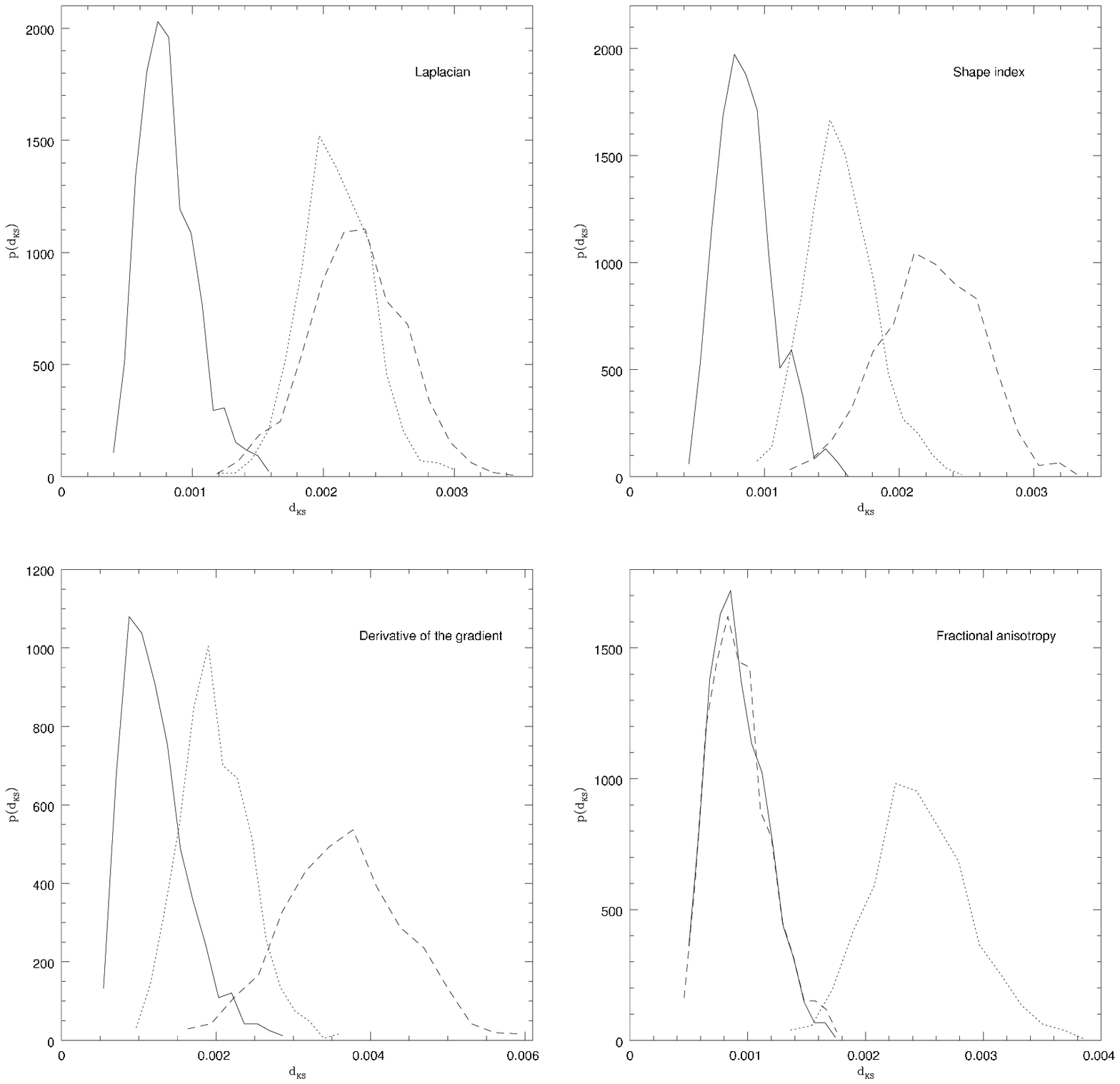}
  \caption{\label{fig:KS_lap_shap_ome_fa_nonoise} Probability distribution 
  of the
  Kolmogorov-Smirnov distances corresponding to the normalised
  Laplacian (top left panel), the shape index (top right panel), the
  derivative of the gradient (bottom left panel) and the fractional
  anisotropy (bottom right panel) of the ideal simulations. Each
  probability distribution is obtained from 1000 simulations for the different
  considered cases: Gaussian (solid line), injected skewness (dashed),
  and injected kurtosis (dotted).}  
\end{center}
\end{figure*}

As a first step, we have applied our Gaussian analysis to the three
types of ideal CMB simulations described in 
table~\ref{tab:Edgeworth_SK_levels}: Gaussian, skewness ($S=0.08$) and
kurtosis ($K=0.4$). The left panel of figure~\ref{fig:KS_temp_nonoise}
shows the probability distribution of KS distances, $p(d_{KS})$,
obtained from the temperature maps of 1000 Gaussian (solid), skewness
(dashed) and kurtosis (dotted) CMB simulations. As expected from the
small values of skewness and kurtosis present in the non-Gaussian
simulations (table~\ref{tab:Edgeworth_SK_levels}), the three curves
completely overlap and thus the temperature distribution cannot
dicriminate between the different cases.
Similarly, we have performed the KS test for each of the normalised
scalars. Figure \ref{fig:KS_lap_shap_ome_fa_nonoise} shows the KS
distance probability distribution of the normalised Laplacian, shape index,
derivative of the gradient and fractional anisotropy. As seen in the
figure, the Laplacian (top left) is able to discriminate very well
between the Gaussian and both non-Gaussian models, since the probability 
distribution 
corresponding to the Gaussian case overlaps only slightly with the ones
of the non-Gaussian simulations. The shape index (top right) and the
derivative of the gradient (bottom left) are able to discriminate
between the Gaussian and skewness simulations but the power of the
test is lower for the kurtosis case. For the fractional
anisotropy the probability distribution of the skewness simulations completely
overlap with the Gaussian one, whereas the case of the kurtosis
simulations is clearly separated.

Table~\ref{tab:powers_scalars} gives the power of the normalised temperature
and the different normalised scalars to discriminate between the ideal
Gaussian and skewness ($S=0.08$) simulations for two signficance levels
($\alpha$ = 0.01, 0.05). The same information is also given
for the ideal kurtosis ($K=0.4$). We remark that the distortion and
the shear have the same discriminating power since they are related
through a strictly monotonous function. Thus we present the results
only for the distortion.
\begin{table*}
 \begin{center} 
  \caption{\label{tab:powers_scalars} Power, $p$, of the KS test at
  two different significance levels ($\alpha=0.05, 0.01$) to
  discriminate between the Gaussian and non-Gaussian simulations for
  the normalised temperature and the normalised scalars. We show the
  results for both the ideal and realistic cases.}

  \begin{tabular}{ c c c c c c c c c }
  \hline
  & \multicolumn{4}{c}{Ideal case} & \multicolumn{4}{c}{Realistic case} \\
  \hline 
  & \multicolumn{2}{c}{S=0.08} & \multicolumn{2}{c}{K=0.40} & \multicolumn{2}{c}{S=0.20} & \multicolumn{2}{c}{K=1.60} \\
  \hline
   & $\alpha=0.05$ & $\alpha=0.01$ & $\alpha=0.05$ & $\alpha=0.01$ & $\alpha=0.05$ & $\alpha=0.01$ & $\alpha=0.05$ & $\alpha=0.01$ \\
 \hline
 Temperature  & 7.5 & 1.9 & 8.4 & 2.3 & 5.3 & 1.3 & 4.6 & 1.5 \\
 Laplacian  & 99.8 & 98.1 & 99.8 & 99.0 & 99.3 & 96.7 & 99.9 & 98.8 \\
 Greatest eigenvalue  & 92.7 & 80.1 & 92.7 & 75.2 & 69.7 & 41.6 & 75.6 & 51.2 \\
 Lowest eigenvalue  & 92.0 & 78.3 & 93.1 & 76.6 & 73.1 & 46.9 & 78.1 & 56.5 \\
 Determinant  & 5.9 & 0.7 & 99.7 & 99.2 & 4.3 & 0.4 & 99.5 & 98.4 \\
 Distortion  & 5.3 & 0.8 & 73.4 & 48.7 & 4.3 & 0.6 & 48.6 & 25.7 \\
 Ellipticity  & 95.2 & 86.9 & 61.1 & 27.8 & 98.4 & 91.9 & 78.0 & 32.9 \\
 Shape index  & 99.5 & 97.9 & 89.3 & 63.2 & 99.1 & 97.1 & 96.1 & 77.3 \\
 Fractional anisotropy  & 6.4 & 2.4 & 99.7 & 98.9 & 6.5 & 1.8 & 100.0 & 100.0 \\ 
 Fractional isotropy  & 99.9 & 98.8 & 91.2 & 73.6 & 99.7 & 97.9 & 93.8 & 72.4 \\
 Gradient  & 8.7 & 2.1 & 66.7 & 43.1 & 4.6 & 1.2 & 72.4 & 51.6 \\
 Derivative of the gradient  & 98.3 & 92.4 & 42.5 & 10.2 & 99.7 & 98.7 & 74.8 & 57.0 \\
 Curvedness  & 6.3 & 1.2 & 99.8 & 99.3 & 5.5 & 0.9 & 95.0 & 85.6 \\
 Gaussian curvature  & 6.8 & 1.7 & 7.4 & 1.9 & 7.9 & 1.8 & 61.2 & 37.5 \\
  \hline
  \end{tabular}
 \end{center}
\end{table*} 

As seen in table~\ref{tab:powers_scalars}, the best scalars at
detecting an excess of skewness are the fractional isotropy, the
Laplacian and the shape index, all of them with $p > 99$ per cent for
$\alpha=0.05$, and also the derivative of the gradient, with a
slightly lower power. Regarding the kurtosis simulations, the scalars
that perform better are the Laplacian, the fractional anisotropy, the
determinant and the curvedness, which all give values of the power $
\ge 99.7$ per cent for $\alpha=0.05$. Note that the performance of the
temperature to discriminate between the Gaussian and non-Gaussian
models is very poor and many of the scalars perform better than the
temperature, especially for the kurtosis case.

Given the very small amount of skewness and kurtosis injected in
the temperature distribution of the non-Gaussian simulations (see
table~\ref{tab:Edgeworth_SK_levels}), it is not surprising that we can
not discriminate between the different considered models using this
quantity. However, we may wonder why some of the scalars are able to
detect the non-Gaussianity. The reason is that this non-Gaussianity
can be amplified in certain scalar maps. To illustrate this point, we
have calculated the mean value and dispersion of the skewness of the
Laplacian for the ideal Gaussian and skewness simulations, finding the
values $(-0.86 \pm 4.03) \times 10^{-3}$, and $(2.89 \pm 0.41) \times
10^{-2}$, respectively. Comparing these values with those given in the
upper part of table~\ref{tab:Edgeworth_SK_levels}, that were obtained
directly from the CMB temperature simulations, it can be seen that the
skewness has been significantly enhanced in the Laplacian
map. Therefore, for the considered case, the non-Gaussianity can be
much more easily detected using the Laplacian than the temperature
map. An analogous argument holds for the kurtosis simulations. In this
case we obtained values for the kurtosis of the Laplacian map of
$(0.01 \pm 7.87) \times 10^{-3}$, and $(7.11 \pm 0.86) \times 10^{-2}$
for the ideal Gaussian and kurtosis simulations, respectively.

From the results of table~\ref{tab:powers_scalars} it also becomes
clear that one scalar can be very good at discriminating one type of
non-Gaussianity whereas its performance can be very poor for a
different type. This points out again the fact that it is not a
trivial task to find a unique Gaussianity test which is optimal in all
cases and motivates the investigation of a relatively large set of
scalars.

We may wonder how our method compares with other Gaussianity tests
proposed in the literature. For instance, techniques based on
spherical wavelets have provided to be a very useful tool to study the
Gaussianity of the CMB
(e.g.~\citealt{barr00},~\citealt{cay01},~\citealt{mar02},~\citealt{vie04}).
In particular,~\cite{mar02} studied and compared the performance of
the spherical Mexican hat wavelet and the Spherical Haar Wavelet using
simulations generetated through the Edgeworth
expansion. Unfortunately, that work is not directly comparable to the
analysis that we present, since there are several differences between
the used methods and simulations, most notably the fact that they use
the Fisher dicriminant to separate between Gaussian and non-Gaussian
distributions and the different levels of injected skewness and
kurtosis. Nonetheless, by looking at their results in table 2 we can
get some insight on the comparative performance of the scalars versus
the spherical wavelets. In particular, it is clear that the best of
our scalars would outperform the spherical Haar wavelet, whereas it
seems that they would provide comparable levels of detection to those
of the spherical Mexican hat wavelet.

\subsection{Realistic case}
\label{sec:noisy}

We have also applied our technique to the realistic CMB simulations
described in section~\ref{sec:simulations}. As in the ideal case, we
have considered three sets of simulations (see table
\ref{tab:Edgeworth_SK_levels}): Gaussian, injected skewness ($S=0.2$)
and injected kurtosis ($K=1.6$). As already explained, we have
increased the levels of non-Gaussianity in these simulations to allow
for a better comparison between the different scalars in the presence
of noise and a mask.

\begin{figure*}
 \begin{center}
  \includegraphics[angle=0,width=15.0cm]{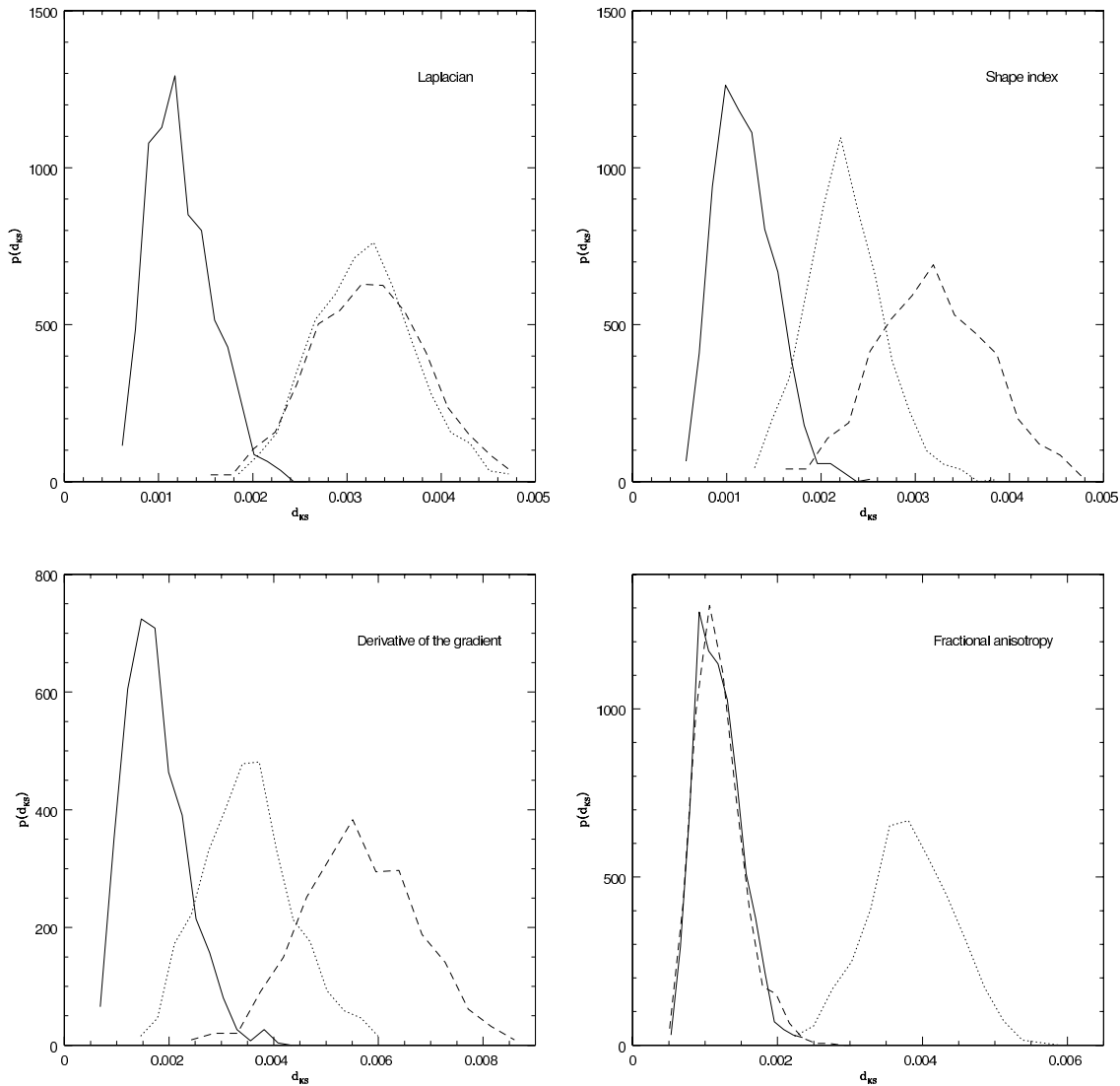}
  \caption{\label{fig:KS_lap_shap_ome_fa_masknoise} Probability distributions 
  of the
  KS distances, corresponding to the Laplacian (top left panel), 
  the
  shape index (top right panel), the derivative of the gradient
  (bottom left panel) and the fractional anisotropy (bottom right
  panel) of the realistic simulations. Each probability distribution is 
  obtained
  from 1000 simulations and the different lines correspond to the
  Gaussian case (solid line), injected skewness (dashed) and
  injected kurtosis (dotted).}  
\end{center}
\end{figure*}

As for the ideal case, we find that the probability distributions of 
KS distances, $p(d_{KS})$, 
corresponding to the Gaussian and non-Gaussian simulations completely
overlap for the normalised temperature (see right panel of
Fig.~\ref{fig:KS_temp_nonoise}). Regarding the scalars, we give in
Fig.~\ref{fig:KS_lap_shap_ome_fa_masknoise} the KS distance
probability distributions of the normalised Laplacian, shape index, 
derivative of the
gradient and fractional anisotropy, for the three sets of realistic
simulations. The results are qualitatively similar to those of the
ideal case: the Laplacian can distinguish reasonably well the Gaussian
case from both types of non-Gaussian simulations, the shape index and
the derivative of the gradient are good at detecting the skewness case
whereas the fractional anisotropy can discriminate only between the
Gaussian and kurtosis simulations.

The results regarding the power of the test at two different levels of
significance ($\alpha=0.01, 0.05$) for the temperature and the
considered scalars are given in table \ref{tab:powers_scalars} for
both the skewness and kurtosis realistic cases. The comparative
performance of the different scalars is very similar to that obtained
for the ideal case, although some scalars seem to be more affected by
the noise than others. In particular, for the skewness case, the
fractional isotropy, the Laplacian and the shape index give again high
values of the power, but they are outperformed by the derivative of
the gradient, that seems more robust under realistic
conditions. Regarding the kurtosis simulations, the fractional
anisotropy, the Laplacian and the determinant perform again very well,
whereas the curvedness is more affected by the presence of noise and
gives a somewhat lower discriminating power.

\section{Conclusions}    

M05 introduced a set of normalised scalars constructed on the
sphere from the covariant derivatives of the temperature field. The
possible use of these quantities for a Gaussian analysis of the CMB
was also proposed. In the present work, following M05, we have
presented a Gaussian analysis for CMB data based on the statitiscal
properties of the normalised scalars. The comparative performance of
the different scalars has been tested using Gaussian and non-Gaussian
simulations generated through the Edgeworth expansion. In particular,
we have used two different types of non-Gaussian simulations: with
injected skewness and with injected kurtosis. In addition we have
considered an ideal case (where noiseless all-sky simulations were
constructed) and a realistic case (containing anisotropic noise and a
mask).

In order to quantify the power of the different scalars to detect
non-Gaussianity we have used the Kolmogorov-Smirnov test. We find
that, in most cases, the scalars amplify the non-Gaussianity present
in the temperature map and produce higher detections than those
obtained directly with the temperature. In particular, the best
scalars to discriminate between the Gaussian and injected skewness
cases are the Laplacian, the fractional isotropy, the shape index and
the derivative of the gradient. For the kurtosis case, the highest
powers are found for the determinant, the fractional anisotropy, the
Laplacian and the curvedness. Note that some scalars can be very good
at detecting one type of non-Gaussianity but perform very poorly for
discriminating the other type.

In future works, we expect to test the discriminating power of these
quantities using physically motivated non-Gaussian models such as the 
quadratic potencial model (\citealt*{lig03}) and also apply them to the 
analysis of the WMAP data.

\section*{Acknowledgments}
The authors thank P. Vielva and H. K. Eriksen for their useful
comments, and also R. Marco for the computational support.  CM thanks
the Spanish Ministerio de Educaci\' on y Ciencia (MEC) for a
predoctoral FPI fellowship.  We acknowledge partial financial support
from the Spanish MEC project ESP2004-07067-C03-01.  This work has
used the software package HEALPix 2.00 (Hierarchical, Equal Area and
iso-latitude pixelization of the sphere,
http://www.eso.org/science/healpix), developed by K.M. G\'orski,
E.F. Hivon, B.D. Wandelt, A.J. Banday, F.K. Hansen and M. Barthelmann.
We acknowledge the use of the software package CMBFAST
(http://www.cmbfast.org) developed by U. Seljak and M. Zaldarriaga.

\appendix

\section{Curvedness}
\label{ap:curvedness_ap} 

In this appendix we describe a new scalar, the curvedness (\citealt{koen90}), 
that was not included in the study of M05. 
The curvedness, $c$, is a scalar defined in terms of the
eigenvalues of the negative Hessian matrix of the field $\lambda_{1}$
and $\lambda_{2}$ by the following expression:
\begin{equation}
c = \frac{1}{2} \sqrt{\lambda_{1}^{2} + \lambda_{2}^{2} }
\label{eq:curvedness_1}
\end{equation}
By construction $c$ is a positive quantity and is closely related to the 
curvature of the field. Note that points with larger values of $\lambda_{1}$ 
or $\lambda_{2}$ (positive or negative) will have larger values of
$c$ whereas it becomes zero for flat areas in the initial field 
(i.e. $\lambda_{1} = \lambda_{2} = 0$). 

Similarly to the other scalars, the normalised curvedness is
constructed replacing $\lambda_1$ and $\lambda_2$ by the normalised
eigenvalues in equation (\ref{eq:curvedness_1}):
\begin{equation}
\tilde{c} = \frac{1}{2} \sqrt{\tilde{\lambda}_1^{2} + \tilde{\lambda}_2^{2} }
\label{eq:curvedness_2}
\end{equation}
When the initial field is Gaussian distributed, the probability distribution 
function of the curvedness is given by:
\begin{equation}
p(\tilde{c})=\frac{16}{\sqrt{\pi}} \tilde{c} e^{-4 \tilde{c}^{2}} D \left(\sqrt{2} \tilde{c} \right)
\label{eq:curvedness_3} 
\end{equation}
which is independent of the power spectrum of the field.
D is the Dawson's function, defined by the integral equation:
\begin{equation}
D(x)= \int_{0}^{x} e^{t^2} dt
\label{eq:dawson}
\end{equation}
The correlations of the normalised curvedness with the other normalised 
scalars are given in table \ref{tab:correl_c_fa_fi}.
\begin{table*}
\begin{center}
    \begin{tabular}{c |  c  c  c  c  c  c  c  c  c  c  c c c}
          & $\tilde{\lambda}_{1}$ & $\tilde{\lambda}_{2}$ &  $\tilde{\lambda}_{+}$ &  $\tilde{d}$  & $\tilde{y}$ &  $\tilde{\lambda}_{-}$ & $\tilde{\iota}$  &  $\tilde{g}$ & $\tilde{D}_{g}$ & $\tilde{\kappa}_{G}$ &  $\tilde{c}$ & $\tilde{f_{a}}$ & $\tilde{f_{i}}$  \\
      \hline
      $\tilde{c}$  & 0.26 & -0.26 & 0 & 0.26 & 0.61 & 0.62 & 0 & 0 & 0 & 0.21 & 1.00 & -0.25 & 0 \\
      $\tilde{f_{a}}$  & 0.23 & -0.23 & 0 & -0.77 & 0.45 & 0.54 & 0 & 0 & 0 & -0.69 & -0.25 & 1.00 & 0 \\
      $\tilde{f_{i}}$  & 0.83 & 0.83 & 0.92 & 0 & 0 & 0 & -0.99 & 0 & -0.53 & 0 & 0 & 0 & 1.00 \\
    \end{tabular}
    \caption{\label{tab:correl_c_fa_fi}
Correlations of $\tilde{c}$, $\tilde{f}_a$ and $\tilde{f}_i$ with the
          rest of normalised scalars (obtained from Gaussian simulations).}
    \end{center}
  \end{table*}

\section{Fractional anisotropy}
\label{ap:frac_aniso_ap} 

In this appendix we describe another interesting scalar, the
fractional anisotropy $f_a$ (\citealt{bass96}). This quantity has
been used in the analysis of medical images, including the
investigation of some neural diseases like astrocytic tumors 
(\citealt{bepp03}) or multiple sclerosis (\citealt{werr99}).

$f_a$ is defined as a function of the eigenvalues $\lambda_1$ and
$\lambda_2$:
\begin{equation}
f_{a} = \frac{1}{\sqrt{2}} \frac{\lambda_{1}-\lambda_{2}}{\sqrt{\lambda_{1}^{2}+\lambda_{2}^{2}}} ~.
\label{eq:frac_aniso_1}
\end{equation}
which is proportional to the ratio of the distortion over the
curvedness.  By construction $f_{a}$ is dimensionless and bounded, $0
< f_{a} < 1$. It encodes information about the asymmetry between the
eigenvalues of the field.

The normalised fractional anisotropy $\tilde{f}_a$ is given by 
\begin{equation}
\tilde{f}_{a} = \frac{1}{\sqrt{2}} \frac{\tilde{\lambda}_{1}-\tilde{\lambda}_{2}}{\sqrt{\tilde{\lambda}_{1}^{2}+\tilde{\lambda}_{2}^{2}}} ~.
\label{eq:frac_aniso_2}
\end{equation}
When the initial field is Gaussian, $\tilde{f}_a$ will follow a
probability density function given by
\begin{equation}
  p(\tilde{f}_{a})=\frac{2\tilde{f}_{a}}{\left( 1-\tilde{f}_{a}^{2}\right)^\frac{1}{2} \left( 1+\tilde{f}_{a}^{2}\right)^\frac{3}{2}}
\label{eq:frac_aniso_3} 
\end{equation}
which, by construction, is independent of the initial power spectrum
of the field.

The correlations of $\tilde{f}_a$ with the other normalised scalars
are shown in table \ref{tab:correl_c_fa_fi}.

\section{Fractional isotropy}
\label{ap:frac_iso_ap} 

The fractional isotropy $f_i$ is proportional to the ratio of the
Laplacian over the curvedness:
\begin{equation}
f_{i} = \frac{1}{\sqrt{2}} \frac{\lambda_{1}+\lambda_{2}}{\sqrt{\lambda_{1}^{2}+\lambda_{2}^{2}}} ~.
\label{eq:frac_iso_1}
\end{equation}
By construction $f_{i}$ is dimensionless and bounded, $-1 < f_{i} < 1$. 

Following the usual construction, the corresponding normalised scalar
$\tilde{f}_i$ is given by:
\begin{equation}
\tilde{f}_{i} = \frac{1}{\sqrt{2}} \frac{\tilde{\lambda}_{1}+\tilde{\lambda}_{2}}{\sqrt{\tilde{\lambda}_{1}^{2}+\tilde{\lambda}_{2}^{2}}} ~.
\label{eq:frac_iso_2}
\end{equation}
For an original Gaussian field $\tilde{f}_i$ follows the pdf:
\begin{equation}
p(\tilde{f}_{i})=\frac{1}{\left(2-\tilde{f}_{i}^{2} \right)^{\frac{3}{2}}} ~.
\label{eq:frac_iso_3} 
\end{equation}
which is independent of the power spectrum of the field.

We include in table \ref{tab:correl_c_fa_fi} the correlations of the
normalised fractional isotropy with the rest of normalised scalars.

\section{Cumulative functions of the normalised scalars}
\label{ap:cum_func_ap}

For some scalars, it is possible to obtain an analytical expression of
its corresponding cumulative function by integrating the pdf's given in table
\ref{tab:norm_scalars}, which are valid for the case of an initial
Gaussian temperature field.

In particular, in table \ref{tab:cum_func}, we give the cumulative
functions of the normalised Laplacian, shear, distortion, ellipticity,
shape index, gradient, fractional anisotropy and fractional isotropy.
\begin{table*}
\begin{center}
    \begin{tabular}{ c  c  c  c }
      \hline
       Normalised scalar & Notation  & Domain&  cumulative \\
      \hline
Laplacian & $\tilde{\lambda}_{+}$& $(-\infty,\infty)$ & 
$F(\tilde{\lambda}_{+})= \frac{1}{2} \left( 1 + erf \left(\frac{\tilde{\lambda}_{+}}{\sqrt{2}} \right) \right) $\\
      \hline
shear & $\tilde{y}$ & $(0,\infty)$ & $F(\tilde{y}) = 1 - e^{-4\tilde{y}} $ \\
      \hline
distortion & $\tilde{\lambda}_{-}$& $(0,\infty)$ & $F(\tilde{\lambda}_{-}) = 1 - e^{-\tilde{\lambda}_{-}^{2}} $ \\
      \hline
ellipticity & $\tilde{e}$ & $(-\infty,\infty)$ & $F(\tilde{e}) = \left\{ \begin{array}{ll} \frac{1}{2} \frac{1}{\sqrt{1+8 \tilde{e}^2}} & \tilde{e} < 0 \\
1 - \left( \frac{1}{2} \frac{1}{\sqrt{1+8 \tilde{e}^2}}\right)  &  \tilde{e} > 0 \\
\end{array} \right .$   \\
      \hline 
shape index & $\tilde{\iota}$& $(-1,1)$ &  $F(\tilde{\iota}) = \left\{ \begin{array}{ll} \frac{1}{2} \left[  1 - \frac{1}{\sqrt{1 + 2 \cot^{2}{\left(\frac{\pi}{2}\tilde{\iota}\right)} }} \right] & \tilde{\iota} < 0 \\
\frac{1}{2} \left[  1 + \frac{1}{\sqrt{1 + 2 \cot^{2}{\left(\frac{\pi}{2}\tilde{\iota}\right)} }} \right] &  \tilde{\iota} > 0 \\
\end{array} \right .$   \\
      \hline
gradient & $\tilde{g}$& $(0,\infty)$ & $F(\tilde{g}) = 1 - e^{-\tilde{g}}$ \\
      \hline
fractional anisotropy & $\tilde{f}_{a}$& $(0,1)$ & $F(\tilde{f}_{a}) = 1 - \sqrt{\frac{1-\tilde{f}_{a}^{2}}{1+\tilde{f}_{a}^{2}}} $\\
      \hline 
fractional isotropy & $\tilde{f}_{i}$ & $(-1,1)$  & $F(\tilde{f}_{i}) = \frac{1}{2} \left( 1 + \frac{\tilde{f}_{i}}{\sqrt{2 - \tilde{f}_{i}^{2}}}\right)$\\
      \hline 
    \end{tabular}
    \caption{\label{tab:cum_func} Cumulative functions of some
normalised scalars, for the case of an initial Gaussian field.}
    \end{center}
  \end{table*}

\end{document}